\documentclass{article}

\usepackage[english]{babel}
\usepackage[square,numbers]{natbib}
\usepackage{authblk}
\usepackage[title]{appendix}
\usepackage[letterpaper,top=2cm,bottom=2cm,left=3cm,right=3cm,marginparwidth=1.75cm]{geometry}

\usepackage{amsmath}
\usepackage{graphicx}
\graphicspath{{images/}}
\usepackage{blindtext}
\usepackage{epstopdf, epsfig}
\usepackage{amsmath}
\usepackage{bm}
\usepackage{enumitem}
\usepackage{varwidth}
\usepackage{placeins}
\usepackage{mathtools}
\usepackage[usenames,dvipsnames]{color}

\newcommand{\pder}[2]{\frac{\partial#1}{\partial#2}}
\renewcommand{\o}[1]{\overline{#1}}
\renewcommand{\l}{\left}
\renewcommand{\r}{\right}

\newcommand{\f}[2]{\frac{#1}{#2}}

\newcommand{\be}{\begin{equation}}
\newcommand{\ee}{\end{equation}}
\newcommand{\pd}[2]{\frac{\partial #1}{\partial #2}}

\newcommand{\bs}{\boldsymbol}

\newcommand{\ucd}{\stackrel{\kern0.1em\triangledown}}
\newcommand{\lcd}{\stackrel{\kern0.1em\triangle}}

\title{Emergent clogging of continuum particle suspensions in constricted channels}

\author[1]{Anushka Herale}
\author[1,*]{Philip Pearce}
\author[2,**]{Duncan Hewitt}

\affil[1]{Department of Mathematics, University College London}
\affil[2]{Department of Applied Mathematics and Theoretical Physics, University of Cambridge}
\affil[*]{philip.pearce@ucl.ac.uk}
\affil[**]{d.r.hewitt@damtp.cam.ac.uk}

\numberwithin{equation}{section}
 \date{}
\begin{document}
\maketitle

\begin{abstract}
    Particle suspensions in confined geometries can become clogged, which can have a catastrophic effect on function in biological and industrial systems. Here, we investigate the macroscopic dynamics of dense suspensions in constricted geometries. We develop a minimal continuum two-phase model that allows for variation in particle volume fraction. The model comprises a ``wet solid'' phase with material properties dependent on the particle volume fraction, and a seepage Darcy flow of fluid through the particles. We find that spatially varying geometry (or material properties) can induce emergent heterogeneity in the particle fraction and trigger the abrupt transition to a high-particle-fraction ``clogged'' state.
\end{abstract}

\FloatBarrier

\section{Introduction}
\label{sec:headings}

Particle suspensions are abundant in industrial and natural systems, from cosmetic and food products to human blood. Dense particle suspensions can exhibit a wide range of complex dynamical behaviour, owing to particle interactions  mediated by the suspending fluid. From a continuum modelling perspective, their effective rheology depends not only on the local shear rate, but also, among other things, on the local particle volume fraction and particle pressure -- a macroscopic representation of isotropic interparticle forces \citep{Guazzelli2018}. One common feature of particle suspensions is that a critical stress must be met to overcome friction and allow particles to rearrange; that is, suspensions have a frictional ``yield stress''. A simple representation of this behaviour within a continuum framework is to impose that below the yield stress, no shear can occur and the solid fraction is equal to a critical ``jamming fraction''. Above the yield stress, the suspension can flow and the volume fraction drops below the jamming fraction. Numerous experimental measurements have captured the divergence of the effective suspension viscosity as the particle fraction approaches the jamming fraction from below \citep{Boyer2011, Zarraga2000,Dbouk_2013, Gallier_2014, Dagois-Bohy_2015}.

In suspensions flowing through confined geometries such as pipes, the concept of clogging can be introduced; a clog is typically considered to be the inability of particles to flow through a confinement \citep{Dressaire2017,Marin2024}. This could be a local and temporary blockage of particle movement because of particles forming intermittent ``bridges'' between walls, or a complete clog in which particle flux is zero everywhere. Even in free-flowing regimes, particle-wall interactions are an important factor in suspension flow. A common assumption is that particles and fluid are unable to move past walls because of frictional effects, leading to a non-uniform shear which is highest at the walls and zero around the centreline; in straight pipes, this leads to radially varying material properties and an unyielded central plug \citep{Lyon1998,Ahnert2019}. In practice, particles are typically able to move past walls to some extent because of a fluid lubrication layer \citep{barnes1995review}, allowing some particle flux even in an unyielded mixture \citep{Szafraniec2022,szafraniec2025suspension}. The unique features of suspensions -- wall slip, yield stress and clogging -- lead to a complex relationship between particle volume fraction and particle flux, and typically the particle flux through channels is maximal at an intermediate volume fraction \citep{Farutin2018}. 

The first aim of our work is to understand the implications of these features at a continuum scale as particle suspensions flow through gradually spatially varying geometries. Local geometrical constrictions have widely been found to promote clogging \citep{Dressaire2017, Dincau2023,Marin2024}. More generally, changes in geometry commonly drive non-intuitive variation in particle volume fraction in dense suspensions, as in the phenomenon of ``self-filtration'' \citep{Haw_2004,Kulkarni_2010}, where driving a high-particle-fraction material through a constriction leads to a lower solid fraction downstream. Theoretical models of suspension flow through constrictions have typically taken an approach that resolves all particles and accounts for their stochastic motion \citep{ Parry2010, Mondal2016, Bacher2017,Marin2024}. However, in general it is not understood how particle dynamics connect to emergent (spatially varying) material properties, which drive continuum-scale suspension flow profiles and particle transport.

The second aim of our work is to explore whether analogies can be drawn between the effects on suspension flow of spatial variations in channel geometry and of spatial variations in particle properties. Biological suspensions containing certain constituents can exhibit such changes at the particle level, with substantial effects on overall suspension flow and function. An example is human blood, which is essentially a suspension containing deformable, fluid-filled capsules: red blood cells (RBCs). The properties of RBCs change as they age, and are also affected during certain drug treatments \citep{Stavroula_2025} and by diseases such as malaria and diabetes. Here, we are interested in spatial variations in RBC properties, which can be drastic in sickle cell disease, a genetic disease that affects the hemoglobin inside RBCs, causing them to stiffen under deoxygenated conditions. Experiments on blood from patients with sickle cell disease have shown that deoxygenation can promote vessel occlusion \citep{Szafraniec2022,Dave2012}, which is linked to poor clinical outcomes. More recently, large spatio-temporal variations in RBC volume fraction (or hematocrit) have been observed to emerge in channels with a deoxygenated region containing stiffened cells~\citep{szafraniec2025suspension} -- it is not clear whether this process is analogous to self-filtration in constricted channels. Phenomenological continuum models have been able to capture vessel occlusion in sickle cell disease by imposing the differential velocity between particles (RBCs) and the suspending fluid \citep{Cohen2013}, but it is not understood how this differential flow emerges mechanistically, i.e. how changes to differential flow follow from changes to particle stiffness.

\begin{figure}
\centering
\includegraphics[width=.9\linewidth]{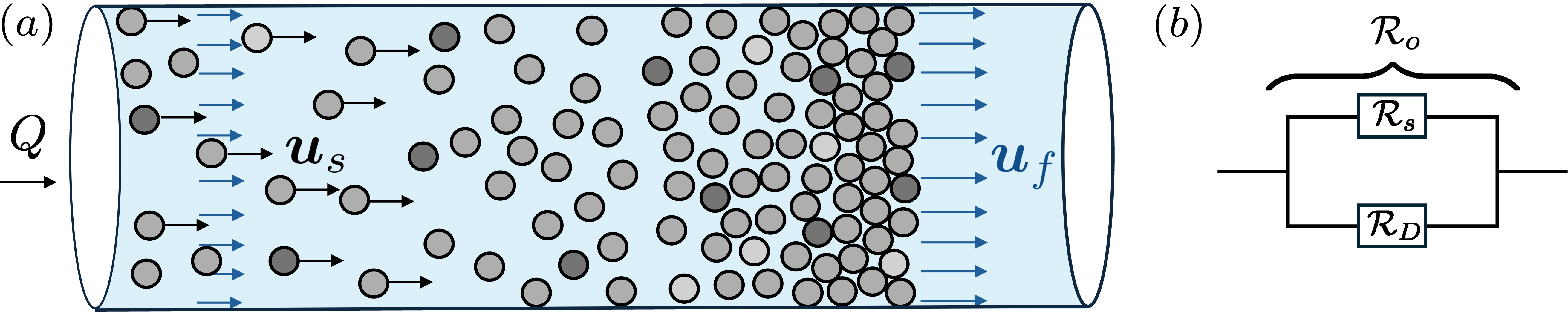}
\caption{(a) A schematic of the ``wet solid'' phase, with velocity $\bs{u}_s$, and the differential ``Darcy'' phase, with velocity $\bs{u}_f$, in pipe flow with a total flow rate $Q$. (b) Equivalent flow resistance schematic with wet solid phase $\mathcal{R}_s$ and differential flow $\mathcal{R}_D$ in parallel.}
\label{fig:schem}
\end{figure}

To achieve these two aims, here we build a minimal, mechanistic continuum model for the flow of a dense particle suspension down a pipe with a constriction in the axial direction. We also allow for varying particle properties via a spatially varying jamming fraction, which has been found to depend on deformability \citep{tapia2024rheology}. For simplicity, we assume no slip at the walls. We allow for the emergence of variations in volume fraction, mediated by differential flow of the suspending fluid past the particles. We find that, for a given constriction or variation in particle properties, flow with a sufficiently low particle fraction can be sustained with a steady, spatially varying profile down the pipe. However, if the intended particle flux crosses a critical threshold, the model predicts the emergence of a high particle-fraction, high resistance, ``clogged'' state, which develops from a free-flowing, unclogged initial state and propagates upstream.
\section{Model}

\subsection{Mass and momentum conservation} 

The overall suspension consists of solid particles, filling a volume fraction $\phi(\bs{x},t)$, and interstitial fluid, occupying a fraction $1-\phi$. We assume the suspension is dense, in the sense that both hydrodynamic and contact forces contribute to the dynamics \citep{Stickel2005,Guazzelli2018}. 
The overall velocity $\bs{U}$ in such a suspension can be broken into the respective local solid and fluid velocities $\bs{u}_s$ and $\bs{u}_f$ as
$
\bs{U} = \phi \bs{u}_s + (1-\phi)\bs{u}_f$ (see schematic in figure \ref{fig:schem}a). Assuming each phase is individually incompressible, conservation of mass in each phase requires that
\be
\pd{\phi}{t} + \bs{\nabla \cdot } (\phi \bs{u}_s )= 0,
\label{mass1}
\qquad
\pd{\phi}{t} - \bs{\nabla \cdot } \left[(1-\phi) \bs{u}_f \right]= 0,
\ee
which combine to yield $\bs{\nabla \cdot U} = 0$. 
Rather than specifying individual, coupled momentum equations for the separate solid and fluid phases \citep{Baumgarten2019}, which can be difficult to compare with rheometric measurements, we instead choose to decompose $\bs{U}$ into the motion of the bulk ``wet solid'' phase, which tracks the solid particles and the interstitial fluid moving at the same speed, and thus has velocity $\phi \bs{u}_s + \left(1-\phi\right) \bs{u}_s  = \bs{u}_s$, and a Darcy seepage velocity, $\bs{u}_D \equiv(1-\phi) (\bs{u_f}-\bs{u}_s)$, which captures the differential transport of fluid through the moving solid particles. Therefore, the part of the fluid motion that moves with the solid particles is tracked within the bulk velocity $\bs{u}_s$, and the part of the fluid motion that moves past the solid is tracked by $\bs{u}_D$. %The benefit of our approach here is that the constitutive laws of both the bulk and Darcy phases are relatively clear and measurable. 
This decomposition gives
\begin{equation}
    \bs{U} %= \bs{u}_s + (1-\phi) (\bs{u_f}-\bs{u}_s) 
= \bs{u}_s + \bs{u}_D.
\end{equation}

Momentum conservation for the overall suspension, neglecting any inertial or gravitational effects, yields
\be
\bs{\nabla \cdot \sigma} = 0,
\label{mom1}
\ee
where $\bs{\sigma}$ is the total stress tensor for the suspension. This stress can be divided into its isotropic and deviatoric parts: the former consists of the fluid pressure $p_f$ and the excess ``particle pressure'' $p_s$ that arises due to the interactions of the solid particles, while the latter we label $\bs\tau$, leading to 
\begin{equation}\bs{\sigma} = - \left(p_f + p_s \right) \bs{I} + \bs{\tau}, \label{stress_def}
\end{equation}
and thus, from (\ref{mom1}), 
\be
\bs{\nabla \cdot \tau} = \bs{\nabla} \left(p_f + p_s \right).
\ee

Based on the premise that the suspension remains in a dense regime, we assume that the stress state of the interstitial fluid is dominated by viscous (Darcy) drag on the particle suspension. As such, deviatoric contributions to the total stress from the motion of the differential fluid are neglected; the differential flow is instead controlled by Darcy's law, which relates fluid pressure gradients to the Darcy seepage velocity,
\be
\bs{u}_D = -\frac{K(\phi)}{\eta_f}\bs{\nabla} p_f,\label{Darcy}
\ee
where $\eta_f$ is the fluid viscosity and $K(\phi)$ describes the permeability of the suspension. 

\subsection{Suspension rheology and constitutive laws}

It has been widely observed that both shear stress and particle pressure in a sheared dense suspension scale linearly with the strain rate $\bs{\dot\gamma} = \bs{\nabla u}_s + \bs{\nabla u}_s^T$ \citep{Guazzelli2018}. Given this observation, we follow \cite{Boyer2011} by assuming a tensorial rheological model for the components of the stress tensor in (\ref{stress_def}) of the form 
\be
p_s = \eta_f\, \eta_n(\phi) \left|\bs{\dot\gamma} \right|,
\qquad
\bs{\tau} = \eta_f\, \eta_s(\phi) \bs{\dot\gamma},
\label{tau1}
\ee
provided $\phi < \phi_m$, where $\phi_m$ is the jamming fraction, with $\bs{\dot\gamma} = \bs{0}$ otherwise. Here $\eta_n(\phi)$ and $\eta_s(\phi)$ are the dimensionless normal and shear viscosity functions, respectively; these are constitutive functions which must diverge as $\phi \to \phi_m$. The tensorial magnitude $\left|\bs{\dot\gamma}\right| = \sqrt{\dot\gamma_{ij}\dot\gamma_{ij}/2}$ denotes the second invariant of that tensor.

This way (\ref{tau1}) of presenting the rheology gives the particle pressure and stress in terms of the solid fraction and strain rate. An entirely equivalent modelling approach is to write the solid fraction and the stress in terms of the dimensionless `viscous number' $J \equiv \eta_f \left|\bs{\dot\gamma}\right|/p_s$, which provides a comparison of characteristic timescales for particle rearrangement and shear (cf. \cite{Boyer2011}), in the form 
\be
\bs{\tau} = \frac{\mu(J)}{J} \bs{\dot\gamma},
\qquad
\phi = \eta_n^{-1}(1/J),
\label{rheol2}
\ee
where $\mu(J) = J \eta_s[\phi(J)]$. The two formulations in (\ref{tau1}) and (\ref{rheol2}) are entirely equivalent. 

Here, we take slightly simplified versions of the constitutive expressions presented by \cite{Boyer2011},
\be
\eta_n(\phi) =   \frac{\phi^2}{(\phi_m - \phi)^2} , \qquad
\eta_s(\phi) =  1 + \mu_1 \eta_n(\phi) ,
\label{eta_ns}
\ee
with both viscosity functions diverging quadratically as the solid fraction approaches its jamming value, $\phi\to \phi_m$, in agreement with numerous experimental measurements \citep{Boyer2011, Zarraga2000,Dbouk_2013, Gallier_2014, Dagois-Bohy_2015}. The viscosity ratio, on the other hand, converges to a finite limiting ``yield friction'' $\eta_s/\eta_n \to \mu_1$ in this limit. 
With these constitutive laws, (\ref{rheol2}) reduces to 
\be
\left. \begin{array}{c}
\bs{\tau} =  \left( \eta_f + \mu_1 p_s /\left|\bs{\dot\gamma}\right|
\right) \bs{\dot\gamma} 
\\
{\phi}/{\phi_m} = 1/\left( 1 + J^{1/2}\right)
\end{array} 
 \right\}
\quad \text{for} \quad \left| \bs{\tau} \right| \geq \mu_1 p_s; 
\qquad
J \equiv \frac{\eta_f \left|\bs{\dot\gamma}\right|}{p_s},
\label{rheol_J}
\ee
with $\bs{\dot\gamma} = \bs{0}$ and $\phi = \phi_m$ otherwise. This representation makes explicit the existence of a ``yield stress'' in the formulation (at $\left|\bs{\tau}\right| = \mu_1 p_s$), which is somewhat obscured in (\ref{tau1}). Below this stress, the material is held at its maximum jamming fraction and cannot deform; above this stress the material deforms and dilates, and the particle fraction is forced to be lower than the maximum jamming fraction. In fact, the expression for $\bs{\tau}$ in (\ref{rheol_J}) resembles the classical Bingham viscoplastic law, with the only difference being that here the yield stress depends on the particle pressure $p_s$. 

Note that taking the full expressions for the constitutive laws in (\ref{eta_ns}) from \cite{Boyer2011} leads to a system that qualitatively resembles (\ref{rheol_J}), sharing the essential features of a pressure-dependent yield stress with viscous flow above yield, but with a slightly more complicated expression for $\bs{\tau}$ that proves rather less analytically tractable in the subsequent analysis. Since our aim here is to provide a general framework to describe suspensions, without placing too much emphasis on specific rheological choices, we retain the simplified form (\ref{rheol_J}) here.

In a similar vein, for the permeability of the packing we adopt the commonly used Carman-Kozeny law \be
K(\phi) = \hat{k} k(\phi) =\frac{\hat{k}(1 - \phi)^3}{\phi^2},
\ee
which comfortingly vanishes when the void space vanishes and diverges as the particle fraction vanishes. The dimensional prefactor $\hat{k}$ is expected to scale with the square of the particle size, and sets the magnitude of the resistance to Darcy flow. This constitutive law is an established model for spherical particles, and in the absence of any more specific particle shape is used illustratively throughout this work. In fact, as we will see, the qualitative conclusions of this work do not depend sensitively on this choice, and the behaviour we observe would also occur even if a constant value was used for the permeability throughout.  

\subsection{Pipe flow, pipe-averaging and lubrication approximation}

For unidirectional, steady flow in a straight pipe of radius $R$ aligned with the $z$ direction, the equations simplify dramatically: only the shear stress $\tau = \tau_{rz}$ and the pressures $p_f$ and $p_s$ play a role, with (\ref{mom1}) reducing to 
\be
\frac{1}{r} \pd{}{r}(  r \tau ) = \pd{}{z} \left( p_s + p_f\right) \equiv -G,\label{mombalance}
\ee
given a total pressure gradient $G$ along the pipe. Integrating yields a linear stress profile across the pipe, $\tau = - {G r}/{2}$.  
%\label{tau1a}
%\ee
Comparison of this linear profile with the rheology in (\ref{rheol2}) shows that the stress must inevitably fall below the yield value $\mu_1 p_s$ at some radius, provided the particle pressure is non-zero, leading to the existence of a ``plugged'' core in the pipe with particle fraction $\phi_m$, as has previously been observed \citep{Ahnert2019}.

Mass conservation (\ref{mass1}) is more usefully presented in its integrated, or pipe-averaged, form,
\be
\pd{\overline{\phi}}{t} + \pd{}{z} \left( \overline{\phi u_s} \right) = 0,
\qquad
\frac{Q}{\pi R^2} = \overline{u_s} + \overline{u_D},
\label{av1}
\ee
where $Q$ is the total flux of the overall suspension down the pipe, $u_s$ and $u_D$ denote the respective flows in the $z$ direction, and the overbar denotes a radial average, so $\overline{f} = (2/R^2) \int_0^R r f \, dr$. The associated total pressure drop along a stretch of pipe of length $\hat{L}$ is $\Delta p = \int_0^{\hat{L}} G \, dz$, 
which motivates the definition of the overall resistance to flow down the pipe,
\be
\mathcal{R}_o = \frac{\Delta p}{Q}.
\label{resis}
\ee
In a similar manner, we can define the individual resistances from the wet solid and Darcy phases, respectively, as $\mathcal{R}_s = \Delta p/ (\pi R^2 \overline{u_s})$ and $\mathcal{R}_D = \Delta p/ (\pi R^2 \overline{u_D})$; these individual resistances effectively act in parallel (see figure~\ref{fig:schem}b) to make up the total resistance, 
\be
\mathcal{R}_o = \frac{\Delta p}{Q} = \left[\frac{1}{\mathcal{R}_s}+\frac{1}{\mathcal{R}_D}\right]^{-1}, \label{resist1}
\ee
which follows from the expression for the total flux in (\ref{av1}). Note that these pipe-averaged equations are one-dimensional, which precludes any issues of ill-posedness that can arise in granular and suspension models \citep{Barker_2015}. 

It is straightforward to generalise these results beyond straight pipes, provided that variations in the pipe geometry occur over a lengthscale $\hat{L}$ that is much larger than the characteristic radial scale $\hat{R}$ of the pipe (i.e. $\partial R/\partial z \ll 1$). This is the realm of classical lubrication theory. 
%The formal approach and scalings are outlined in Appendix \eqref{AppendixND}; 
After scaling radial lengths with $\hat{R}$, axial lengths with $\hat{L}$, velocities with an as-yet undetermined scale $V$, stresses and particle pressures with $\eta_f V/\hat{R}$  and fluid pressure with $\eta_f V \hat{L}/\hat{R}^2$, and working under the assumption that $\hat{R} \ll \hat{L}$, we arrive at the following equations, written now in terms of dimensionless variables, 
\be
G = -\pd{p_f}{z}, \qquad 
\tau = - \frac{G r}{2}, \qquad
u_D = -Da \, k(\phi) \pd{p_f}{z},
\label{lub1}
\ee
to leading order in $\hat{R}/\hat{L}$, while the rheology reduces to 
\be
p_s = \eta_n(\phi) \left| \pd{u_s}{r} \right|,
\qquad
\tau = \eta_s(\phi) \pd{u_s}{r},\label{pp_tau}
\ee
provided $\phi < \phi_m$. The Darcy number 
\begin{equation}
    Da = \frac{\hat{k}}{\hat{R}^2},
\end{equation}
determines the role of the permeability, and compares the square of the typical particle size, $\hat{k}^{1/2}$, with the square of the pipe radius, $\hat{R}$; it should, therefore, be significantly smaller than unity for a continuum description of the suspension to be reasonable.

The pipe-averaged equations (\ref{av1}), written in terms of dimensionless variables, become
\be
\pd{\overline{\phi}}{t} + \frac{1}{R^2}\pd{}{z} \left( R^2 \overline{\phi u_s} \right) = 0,
\qquad
\frac{Q}{\pi R^2} = \overline{u_s} + \overline{u_D},
\label{av2}
\ee
where $R(z)$ is the dimensionless pipe radius. The definitions of the resistances remain as in (\ref{resist1}). Note that one can set the undetermined velocity scale $V$ to scale out either the total flux $Q$ or the total pressure drop $\Delta p$ from the problem, but in either case the resistance in (\ref{resist1}), which is the ratio of the two, is unchanged. From this point on, we focus on the case of an imposed total flux, and thus choose the velocity scale $V$ so as to scale the dimensionless flux $Q$ to unity (i.e. $V = Q/\pi \hat{R}^2$), although we briefly revisit this question of whether the flux or pressure-drop is imposed in Appendix~\ref{ap_press}. The solid flux that appears in (\ref{av2}) will prove an important quantity, and so we label it as 
\begin{equation}
    \mathcal{F} = R^2 \overline{\phi u_s}.
    \label{Fdef}
\end{equation}
Note that, given the choice of scaling the dimensionless flux to unity, $\mathcal{F}$ here represents the fraction of the total volume flux that is made up of solid. 

As a side note, a key simplification arising from these lubrication scalings is the first equation in (\ref{lub1}): the suspension rheology (\ref{tau1}) enforces that the particle pressure $p_s$ must scale with the shear stress $\tau$, but the usual lubrication scalings enforce that the fluid pressure $p_f$ is asymptotically larger (by a factor $\hat{L}/\hat{R}$). As such, particle pressure gradients are negligible compared with fluid pressure gradients, and so can be ignored in the overall pressure gradient $G$: the suspension is pushed along the channel by fluid (pore) pressure gradients to leading order. As a result of this difference in scaling for the two pressures, (\ref{av2}) becomes a hyperbolic equation for $\overline{\phi}$. In principle, one could include the asymptotically small particle-pressure gradients, which would add a weak non-local and non-hyperbolic character to the model that should slightly smooth over sharp fronts and shocks when they arise.

In summary, the model comprises the evolution equation for $\overline{\phi}(z,t)$ in (\ref{av2}), which, in turn, is determined by knowledge of the pipe-averaged quantities $\overline{\phi}$, $\overline{u_s}$, $\overline{u_D}$ and $\overline{\phi u_s}$. These can be deduced from the local force balances across the pipe and the suspension rheology (\ref{lub1})--(\ref{pp_tau}), as outlined at the start of the next section. The parameters of the model are the Darcy number $Da$, which controls the permeability of the suspension, and the maximum solid fraction $\phi_m$ of the suspension, together with the other rheological constants and any imposed variation in the pipe radius $R(z)$ (in \S\ref{emerge} we briefly consider other forms of variation along the pipe). Throughout this work, we take the limiting friction coefficient to be $\mu_1 = 0.3$, following \cite{Boyer2011}.

We take a fixed total flux, which in dimensionless form is scaled to give $Q = 1$, and consider a fixed inlet solid fraction $\overline{\phi}_{in} = \overline{\phi}(z=0,t=0)$ at the start of the pipe. In a steady state, the model simply reduces to two algebraic statements of flux conservation: the total flux $Q \equiv 1$ and the relative solid flux $\mathcal{F}$ (\ref{Fdef}) are both fixed. The time-dependent model (\ref{av2}), on the other hand, is a hyperbolic first-order partial differential equation, details of which are discussed in \S\ref{emerge} below and Appendix~\ref{AppendixNum}.

\begin{figure}
\centering
\includegraphics[width = .9\linewidth]{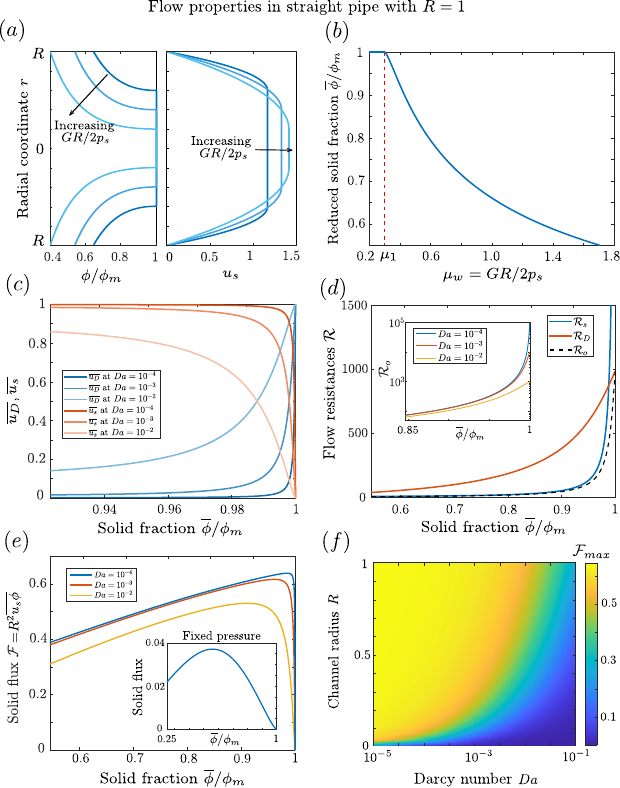}
\caption{Flow properties in a straight pipe, with $R=1$ unless otherwise specified. $(a)$ Varying particle fraction and solid velocity profiles for increasing shear to normal stress ratios, $\mu_w = G R/2 p_s = 0.5,0.75,1.5$, and $(b)$ scaled average solid fraction $\o{\phi}/\phi_m$ as a function of $\mu_w=G R/p_s$. $(c)$ Average wet solid and Darcy phase velocities varying with $\o{\phi}/\phi_m$ for different $Da$. $(d)$ Flow resistances (\ref{resis}--\ref{resist1}) for $Da = 10^{-2}$ varying with $\o{\phi}/\phi_m$; the inset shows how the overall resistance for $\o{\phi}$ near to $\phi_m$ increases as $Da$ is reduced. $(e)$ Solid flux showing a maximum at a value of $\o{\phi} < \phi_m$, for different $Da$ ; the inset shows the scaled solid flux $\mathcal{F}/\mathcal{R}_o$, which is independent of $Da$, if the overall pressure gradient is fixed, rather than the overall flux. $(f)$ The maximum achievable solid flux, which decreases as the pipe radius decreases and as the Darcy number (permeability) increases.}
\label{fig:2_II}
\end{figure}

 \section {Results}

\subsection{Flow and volume fraction profiles in a straight pipe} 

From (\ref{lub1})--(\ref{pp_tau}), we see that the ratio of the shear stress and the particle pressure yields a function of $\phi$ alone,
%\begin{gather}
   $\left|{\tau}/{p_s}\right| = {\eta_s(\phi)}/{\eta_n(\phi)} = {G r}/{ 2 p_s}$, %\label{tau/pp}
%\end{gather}
provided $\phi<\phi_m$. Given the specific rheological functions in (\ref{eta_ns}), this equation reduces to 
\begin{gather}
    \phi(r,t) = \f{\phi_m\sqrt{R}}{\sqrt{R} + \sqrt{\mu_w \l(r - R \mu_1/\mu_w\r)}} \quad \text{for} \quad r> r_c,   \label{phiprofile}
\end{gather}
where 
\be
\mu_w(t) \equiv \f{GR}{2 p_s},
\qquad \text{and} \qquad
r_c(t) \equiv \frac{R \mu_1}{\mu_w} = \frac{2 p_s \mu_1}{G},
\label{rc}
\ee
with $\phi(r<r_c) = \phi_m$. Here, the grouping $\mu_w$ combines the local pipe radius, pressure gradient and particle pressure, and can be thought of as an effective friction coefficient for the suspension at the pipe walls.
The critical ``yield'' radius $r_c$ separates a central core of material with $\phi = \phi_m$ from a flowing region with $\phi<\phi_m$ given by (\ref{phiprofile}). Example plots of the profiles of $\phi$ predicted by (\ref{phiprofile}) are shown in figure~\ref{fig:2_II}(a).

Given $\phi$, the associated velocity profile $u_s$ follows from the expression for $\tau = -Gr/2$ in \eqref{lub1}, and thus the expression for $\partial u_s /\partial r$ in (\ref{pp_tau}). We assume here that the suspension experiences a no-slip condition ($u_s = 0$) at the walls of the pipe $r=R$, and as such can integrate  $\partial u_s /\partial r$ to give the yield-stress like flow profile
\begin{gather}
    u_s(r,t) = \begin{dcases}
  \f{G}{4}(R - r_c)^2 \hspace{0.3cm} &\text{for  $r < r_c$}\\
  \f{G}{4}\l[(R - r_c)^2 - (r - r_c)^2\r] \hspace{0.3cm} &\text{for  $r \geq r_c $},
             \end{dcases}\label{velprof}  
\end{gather}
(the same expression can equivalently be reached by directly integrating the expression for the stress in (\ref{rheol_J}) using (\ref{pp_tau})).  Equation (\ref{velprof}) describes a central plug of unyielded material (corresponding to where $\phi=\phi_m$), bordered by sheared regions, as illustrated in figure \ref{fig:2_II}(a).  This qualitative flow structure for dense suspensions in a channel or pipe is well known \citep{Lyon1998,Koh1994,Phillips1992}. 

The cross-pipe profiles for $\phi$ and $u_s$ provide all the necessary information to determine the pipe-averaged quantities in (\ref{av2}). These integrals can be computed analytically, but are algebraically unpleasant: details of this integration and the form of the resultant average quantites are given in Appendix {\ref{Appendix1}. Note that the structure of the flow across the pipe depends only on the grouping $\mu_w = G R/2p_s$, which sets the radius of the central plug region (\ref{rc}), rather than on the pressure gradient or particle pressure alone. It is, therefore, this grouping $\mu_w$ that controls whether the suspension is able to flow or not: the suspension formally clogs (i.e. the solid flux goes to zero) only if the central plug fills the pipe, $r_c \to R$, in which limit $\overline{\phi} \to \phi_m$. 

 Figure~\ref{fig:2_II}(b) shows the dependence of the average solid fraction $\overline{\phi}$ on $\mu_w$, which follows from the radial integral of (\ref{phiprofile}). This plot illustrates directly the above point: $\overline{\phi} \to \phi_m$ when $G R \leq 2\mu_1 p_s$, as the central plug fills the entire pipe. The associated solid flux will also vanish in this limit, owing to the no slip condition on the walls.

\subsection{Role of the Darcy phase: flow resistance and maximum particle flux} 

\label{darcy_f_sec}

The results so far are independent of the differential (Darcy) flow of fluid through the suspension. Its role is to modulate the total flux $Q$, through (\ref{av2}), which provides another constraint linking $G$ and $p_s$, and thus determines the behaviour of the system explicitly (see Appendix \ref{Appendix1} for the explicit expressions). Figure~\ref{fig:2_II}(c) shows the individual mean ``wet solid'' and Darcy velocities as a function of the mean solid fraction, given that the total flux is fixed at $Q=1$. It is clear that the solid velocity dominates the flux until the solid fraction approaches its maximum packing. In this limit, the flow resistance from the wet solid phase is so great that fluid is forced through the packing instead: both the pressure gradient $G$ and the particle pressure $p_s$ grow significantly over a narrow range of $\overline{\phi}$ in order to maintain the desired solid fraction whilst providing the necessary total flux. As such, the Darcy velocity grows near $\overline{\phi} = \phi_m$ as the solid flux drops towards zero in this limit. The Darcy number $Da$ controls the permeability of the packing, and thus the resistance to Darcy flow; hence the region in which the Darcy flow becomes dominant is increasingly localised to be close to $\phi_m$ as $Da$ is reduced (figure~\ref{fig:2_II}c). Note that the pressure gradient $G$ drives both the bulk wet solid motion and the Darcy seepage flow along the pipe, and thus for any $G$ there is always some Darcy flow, but its contribution is generally negligible unless the solid fraction is near its maximum value (as can be seen in figure~\ref{fig:2_II}c).

The competition between ``wet solid'' flow and Darcy flow is perhaps better illustrated in figure~\ref{fig:2_II}(d), which shows the individual resistances for each phase and the overall flow resistance, as defined in (\ref{resist1}). The ``wet solid'' resistance is generally much lower than the Darcy resistance, but it increases dramatically as $\overline{\phi} \to \phi_m$. Because the two resistances effectively act in parallel (figure~\ref{fig:schem}(b) and (\ref{resist1})), the overall resistance (dashed line) becomes limited by the Darcy resistance in that limit instead. 

As a consequence of this behaviour, the solid flux down the pipe (as a fraction of the overall flux) $\mathcal{F} = R^2 \overline{\phi u_s}$ does not simply increase monotonically as $\overline{\phi}$ is increased (figure~\ref{fig:2_II}e): it has a maximum at a solid fraction lower than $\phi_m$, before reducing again for high solid fraction and vanishing in the limit $\overline{\phi} \to \phi_m$, when the particles becomes stationary (because there is no slip of particles on the walls in our model). This means that any given solid flux (below the maximum) can be attained at two different values of the solid fraction: a low $\overline{\phi}$, with a correspondingly high solid velocity and relatively low resistance; or a high $\overline{\phi}$, with a correspondingly low solid velocity and high resistance. Again, lowering the Darcy number (permeability) increases the resistance to porous flow and tends to squash the `high-$\overline{\phi}$' branch of the flux closer to $\overline{\phi}= \phi_m$ (figure~\ref{fig:2_II}e). Note that the same qualitative behaviour of a non-monotonic solid flux is also seen if the pressure gradient $G$ is held fixed, rather than the total flux (inset to figure~\ref{fig:2_II}e).

More generally, these solid flux profiles $\mathcal{F}(\overline{\phi}; Da)$ also vary with the pipe radius (figure~\ref{fig:2_II}f), with smaller radii offering a lower maximum solid flux $\mathcal{F}_{max}$. The reason for this reduction in solid flux is that the relative contribution of the Darcy flux to the total flux increases in thinner pipes: given a pipe with cross-sectional area $A\sim R^2$, the solid flux scales with $A G R^2 \sim  G R^4$, because $u_s \sim GR^2$ (from Eq.~\ref{velprof}), whereas the Darcy velocity is constant across the pipe and the Darcy flux scales like $\sim AG \sim GR^2$ (see also Appendix~\ref{Appendix1}). This reduction in the attainable solid flux with radius will prove a crucial detail for interpreting the behaviour in constricted pipes where $R$, and thus all the quantities calculated here, can vary along the pipe.

\begin{figure}
\centering
\includegraphics[width = \linewidth]{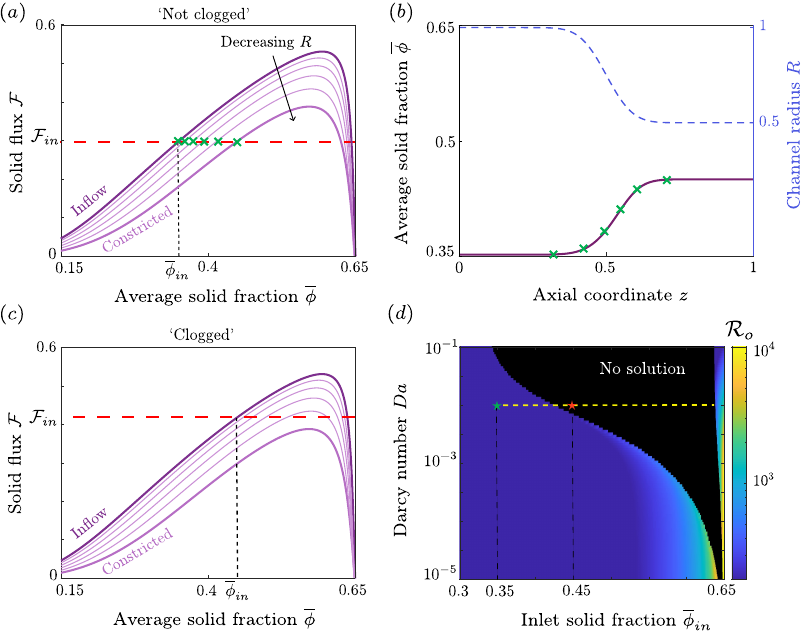}
\caption{ Illustration of `non-clogging' and a `clogging' constrictions. $(a)$ Solid flux $\mathcal{F}(\o{\phi})$ for $\phi_m = 0.65$, $Da = 10^{-2}$ and different values of the radius $R$ varying between $R=1$ and $R = R_{min} = 0.5$. The red dashed line shows $\mathcal{F}_{in}$ for $\o{\phi}_{in} = 0.35$. $(b)$ Steady solution $\o{\phi}(z)$ for an imposed radial constriction between $R=1$ and $R= R_{min} = 0.5$ as shown by the blue dashed line; the green crosses correspond to the crosses in $(a)$. $(c)$ The same flux curves as in $(a)$, but now showing $\mathcal{F}_{in}$ for a slightly larger inlet solid fraction $\o{\phi}_{in} = 0.45$; there is no steady solution with this value of $\o{\phi}_{in}$ because the downstream constriction cannot sustain this flux. $(d)$ Heatmap of overall resistance against $\o{\phi}_{in}$ and $Da$ for the same constricted system; the dashed line represents $Da = 10^{-2}$ and the two stars correspond to the cases illustrated in $(a)$ and $(c)$.}
\label{fig:3}
\end{figure}

\subsection{Flow and particle flux in constricted pipes}\label{constriction_section}
We now consider pipes with an imposed radial constriction $R(z)$ (figure~\ref{fig:3}), in which the radius decreases smoothly from $R=1$  to $R = R_{min}$ as shown in figure~\ref{fig:3}(b). For all cases shown here, we adopt a profile 
\begin{equation}
R(z) = \frac{ \left(1+R_{min} \right) - \left(1-R_{min}\right) \text{erf}\left[\left(z-0.5\right)10\right]}{2}.
\label{radial_z}
\end{equation}
Provided that gradients in the true (dimensional) pipe radius are sufficiently small, the evolution of solid fraction is described by (\ref{av2}), with the local flux at any location $R(z)$ given by the flux for a straight pipe of that radius. As such, the steady-state flow is determined simply by a statement of flux conservation: the solid flux through each section of the pipe must be equal, which should allow for determination of the associated average profile $\o{\phi}(z)$. 

Figure~\ref{fig:3}(a,b) shows such a steady solution. Given a value of the inlet solid fraction $\o{\phi}_{in}$, the local solid fraction along the pipe $\o{\phi}(z)$ follows directly from the expression for the solid flux $\mathcal{F}(\o{\phi},R)$ evaluated at each local radius $R(z)$ (figure~\ref{fig:3}a). Since the solid flux is conserved, the variation in the volume fraction along the pipe can simply be read off from the intersection of the different solid flux curves for each radius with the incident solid flux $\mathcal{F}_{in}$ (green crosses in figure~\ref{fig:3}a), leading to a profile $\o{\phi}(z)$ that increases slightly in the constricted part of the channel (figure~\ref{fig:3}b) -- this effect has been observed qualitatively in particle-scale simulations~\citep{Bacher2017}.

However, this solution construction is not always possible. For sufficiently high inlet volume fractions $\o{\phi}_{in}$, or sufficiently large constrictions, it is no longer possible to determine the local solid fraction using the incident solid flux and the local radius in this manner. This is the case when the solid-flux curve for the most constricted part of the pipe is everywhere lower than the solid flux set by the inlet solid fraction (as illustrated in figure~\ref{fig:3}c), and so there is no steady solution to (\ref{av2}) that conserves the imposed inlet solid flux in that part of the channel. %For the range of parameters in this regime, we are unable to find a steady solution to (\ref{av2}) using a numerical approach (figure~\ref{fig:3}d).

 A full parameter sweep illustrates that as $\o{\phi}_{in}$ increases, the overall pipe resistance increases until the ``no solution'' region of parameter space is reached; figure~\ref{fig:3}(d) shows this parameter space for a fixed radial constriction with $R_{min} = 0.5$. Note that for sufficiently high $\o{\phi}_{in}$, the associated inlet solid flux $\mathcal{F}_{in}$ reduces (see flux curves in figure~\ref{fig:3}a) and a steady solution can again be found. In this region of parameter space, $\o{\phi}$ is always on the upper branch of the flux curve: the solid fraction now decreases through the constriction, rather than increases, and is everywhere fairly close to the jamming fraction, such that the overall flow resistance can be very large (figure~\ref{fig:3}d).

\subsection{Emergent spatio-temporal heterogeneity in constricted pipes}
\label{emerge}

To understand the dynamics of the system in the regions of parameter space that do not appear to exhibit a steady solution (black region in figure~\ref{fig:3}d), we solve the time-dependent problem \eqref{av2} using the Clawpack finite-volume method for hyperbolic problems \citep{clawpack} (see Appendix~\ref{AppendixNum} for further details). We use an initial condition given by the steady state for $\o{\phi}_{in} = 0.2$, and consider a step increase in the injected solid fraction at $z=0$ to a new (higher) value of $\o{\phi}_{in}(t=0)$. We subsequently enforce reflection conditions ($\partial \o{\phi}/\partial z = 0$) at $z=0$, rather than holding the inlet particle fraction fixed, in order to allow for variations in $\o{\phi}$ to emerge throughout the entire pipe.

\begin{figure}
\centering
\includegraphics[width = \linewidth]{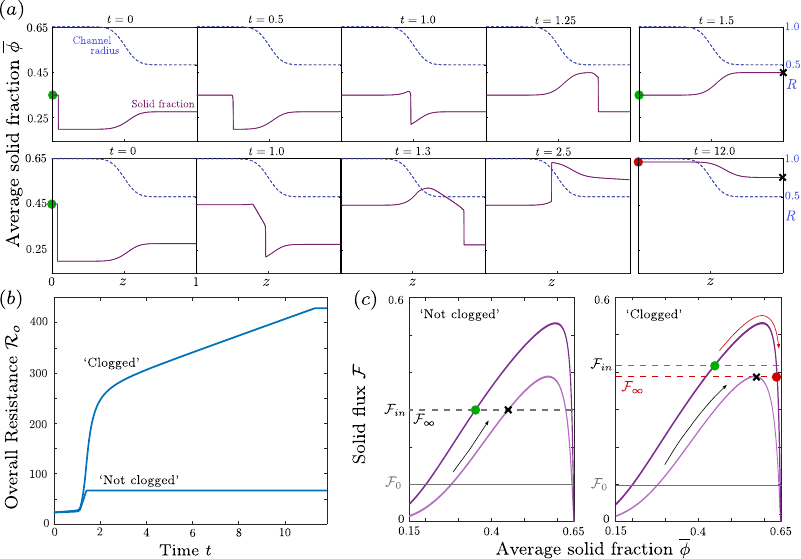}
\caption{Transient evolution of constricted flow. $(a)$ Evolution from a low solid-fraction steady state, with $\o{\phi}_{in} = 0.2$, when  $\o{\phi}_{in}(t=0) = 0.35$ (upper) and $\o{\phi}_{in}(t=0) = 0.45$ (lower), for $\phi_m = 0.65$, $Da = 10^{-2}$ and $R_{min} = 0.5$ 
 (with $0\leq z\leq 1$ for all panels). These solutions correspond to a ``not clogged'' (upper) and emergent ``clogged'' (lower) state, respectively, with symbols corresponding to those marked in $(c)$. $(b)$ Overall resistance $\mathcal{R}_o$ over time for each case. $(c)$ The solid flux $\mathcal{F}$ for the widest and narrowest parts of the pipe, for the ``not-clogged'' (left) and ``clogged'' (right) examples in panel (a). The pre-existing steady state has flux $\mathcal{F}_0$ that is initially increased at $z=0$ to $\mathcal{F}_{in}$ (green dot). In the right-hand case, the downstream flux (black cross) is too low, and the upstream solid fraction is forced to increase towards the red dot, reducing the eventual solid flux $\mathcal{F}_{\infty}$, as outlined in the main text. }
\label{fig:4}
\end{figure}

\begin{figure}
\centering
\includegraphics[width = \linewidth]{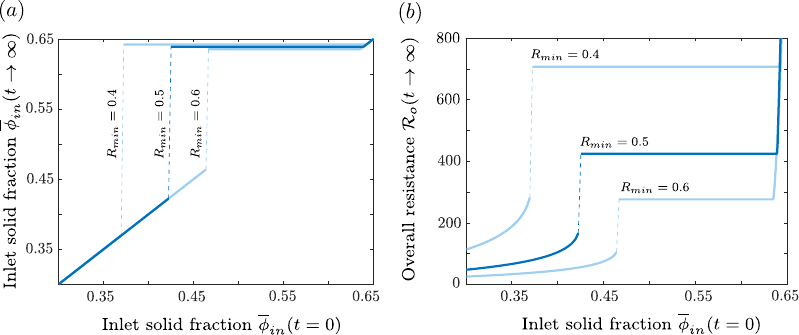}
\caption{The final steady-state $(a)$ upstream (inlet) solid fraction and $(b)$ total resistance as a function of the initial inlet solid fraction, for a constricted pipe with minimum radius as marked, $Da = 10^{-2}$ and $\phi_m = 0.65$. The central line, which matches the parameters in figure \ref{fig:4}, corresponds to traversing the yellow dashed line in the phase plot in figure~\ref{fig:3}(d). The discontinuities in each case represent the development of the upstream-propagating `clogged' state.}
\label{fig:Resist_new}
\end{figure}

We obtain temporal solutions in regions of parameter space both with and without a steady solution. For the former, the system reaches the steady state described above in which the solid fraction increases slightly within the constricted part of the channel (figure~\ref{fig:4}a upper panels). In the ``no solution'' region of parameter space, however, while the solid fraction again increases downstream, it now starts to `back-up' solid, and forms a shock just upstream of the constriction. The shock then travels upstream all the way back to the inlet (figure~\ref{fig:4}a lower panels), raising the solid fraction significantly everywhere upstream.  Note that the existence of a shock -- that is, an axial discontinuity in solid fraction -- locally violates our lubrication approximation; in reality we might expect this backwards-propagating region to be smoothed out over a dimensional length comparable to the pipe radius, with locally more complex and non-unidirectional dynamics. The associated overall resistance steadily increases as the shock moves backwards, extending the region where the solid fraction - and thus the flow resistance - is high (figure~\ref{fig:4}a,b). Thus, as the shock passes through the inlet the solid fraction $\o{\phi}_{in}$ increases to a much higher value, which we designate $\o{\phi}_{in}(t \to \infty)$, and a new steady state is reached.

We refer to this new state -- shown in figure ~\ref{fig:4}(a-c) -- as ``clogged''. This is for two reasons: first, because of its very high solid fraction and overall resistance, and second, because of the way it emerges discontinuously to fill the pipe from an apparently free-flowing, moderate solid-fraction state. In this manner, the emergence of the ``clogged'' state provides a downstream control on upstream properties (via the backwards propagating shock), in direct contrast to the ``non-clogged'' state, in which the imposed upstream solid fraction $\phi_{in}$ sets the profile down the pipe. For clarity, we note that in our model, the solid flux is not reduced to zero in this ``clogged'' state, which has $\o{\phi}<\phi_m$ everywhere. However, the resistance increases so dramatically (figure~\ref{fig:4}b) that an experimental apparatus imposing a fixed overall flux (as in our theory) might be expected to fail in this regime. Alternatively, the apparatus may be unable to adjust to a new, higher, inlet particle fraction, as occurs in our model, which would also lead to failure. If, instead, the pressure drop down the pipe were fixed, the overall flux would decrease significantly in the ``clogged'' state, as briefly discussed in Appendix~\ref{ap_press}. 

The transient solutions indicate how to find this ``clogged'' steady state directly. If the imposed solid flux is too large for the constricted pipe (i.e. in the `no-solution' region of phase space; figure~\ref{fig:3}d) then the flux is lowered until a solution can be found. Physically, this occurs by `backing up' the excess solid upstream of the constriction, which pushes the upstream solid fraction up onto the high-$\o{\phi}$ branch of its flux curve. This process is illustrated in figure~\ref{fig:4}(c), which shows how the solid flux evolves in the pipe for each of the two cases considered in panel (a). In each case, the initial steady state carries a flux $\mathcal{F}_0$, and at $t=0$ this is increased at $z=0$ to $\mathcal{F}_{in}$. In the ``not clogged'' case, this new flux can be sustained all along the pipe, and the suspension simply evolves towards the associated steady state; there is no subsequent change in the solid flux and $\mathcal{F}_{\infty} = \mathcal{F}(t\to \infty)$ is simply $\mathcal{F}_{in}$. In the ``clogged'' case, however, the flux is limited by the downstream maximum: incoming solid builds up upstream of the constriction, and the only resolution to this mismatch is for the upstream solid fraction to increase to the point where the flux is reduced to match the downstream limiting value. Thus $\mathcal{F}_{\infty}$ is lower than $\mathcal{F}_{in}$, as shown in figure~\ref{fig:4}(c).

The eventual emergent state in this case can therefore be directly constructed by finding the limiting downstream flux, and following the high-$\o{\phi}$ flux branch upstream from this value.  Equivalently, this steady state corresponds to the point where a horizontal line in a phase-space plot like figure~\ref{fig:3}(d) meets the right-hand boundary of the ``no solution'' region.  

We find that this construction holds for any initial condition within the ``no solution'' region. Figure \ref{fig:Resist_new}(a) shows how the eventual solid fraction at the start of the pipe, $\o{\phi}_{in}(t \to \infty)$, varies with the initial solid fraction at the start of the pipe. These quantities are equal when the solid fractions are sufficiently small, but above some critical $\o{\phi}_{in}$ the fluxes down the pipe can no longer balance, as outlined above, and the system is forced into this new state with a much higher solid fraction throughout the pipe. The associated resistance is also significantly increased, as shown in figure~\ref{fig:Resist_new}(b). If the size of the constriction is increased (i.e. $R_{min}$ decreased), the transition to this emergent ``clogged'' state is more more dramatic, and occurs at a lower inlet solid fraction (figure~\ref{fig:Resist_new}). Note that the both the final inlet solid fraction and the resistance increase again for $\phi$ very close to $\phi_m$; this region of parameter space corresponds to the steady states that lie entirely on the high-$\o{\phi}$ branch of the flux curves, as discussed at the end of the previous section and illustrated by the band of steady solutions close to $\phi_m$ in figure~\ref{fig:3}(d).

\subsection{Emergent heterogeneity in straight pipes with variation in particle properties}

Motivated by recent experiments on blood flow from patients with sickle cell disease~\citep{Szafraniec2022,szafraniec2025suspension}, in which red blood cells stiffen under deoxygenated conditions, we also briefly consider suspension flow in straight pipes with spatial variation in particle properties. To represent such variations in the simplest way possible, we allow the maximum particle fraction $\phi_m(z)$ to decrease smoothly from high to low through the pipe (figure~\ref{fig:5}a), qualitatively capturing the reduction in jamming fraction for suspensions containing rigid particles in comparison to more deformable particles \citep{tapia2024rheology}. 
In so doing, our aim is to explore whether analogies can be drawn between variation in particle properties and the geometrical constrictions studied above. We also aim to provide at least a basic qualitative explanation of experimental observations of large-scale spatial variations in hematocrit (i.e. particle fraction) for blood flow in channels with a downstream deoxygenated region that contains stiffened red blood cells \citep{szafraniec2025suspension}}. %

We find that steady solutions of \eqref{av2} with varying $\phi_m(z)$ exhibit a bifurcation analogous to the one observed for channel constrictions: if the incident particle fraction  $\o{\phi}_{in}(t = 0)$ is too large relative to the downstream maximum particle fraction, then the incident solid flux $\mathcal{F}_{in}$ cannot be maintained, and a shock forms and propagates upstream. This behaviour is illustrated in figure \ref{fig:5}(a), which shows results from time-dependent simulations in which an initially uniform solid fraction is exposed at $t=0$ to a prescribed reduction in the jamming fraction $\phi_m$ along the pipe. If the initial solid fraction is sufficiently low (upper row), the suspension evolves smoothly to a steady state that carries the same solid flux $\mathcal{F}_{in}$ as before. However, larger initial solid fractions (lower row) instead result in a shock developing and spreading upstream, leading to a final state with a much higher solid fraction, that carries a lower solid flux $\mathcal{F}_\infty$. As in the case of a constriction, the associated overall flow resistance increases markedly in this ``clogged'' state (figure \ref{fig:5}b). The final states can again be constructed simply by consideration of the solid fluxes, as illustrated in figure \ref{fig:5}(c). 

These results suggest that the spatial variations observed in previous experiments on blood flow \citep{szafraniec2025suspension} may be explained by a reduction in maximal red blood cell flux as cells become stiffened under deoxygenation. More broadly, the results demonstrate that our predictions of emergent heterogeneity in solid fraction apply to a broad range of synthetic and biological particle suspension systems in which either geometry or particle properties vary in space.

\begin{figure}
\centering
\includegraphics[width = \linewidth]{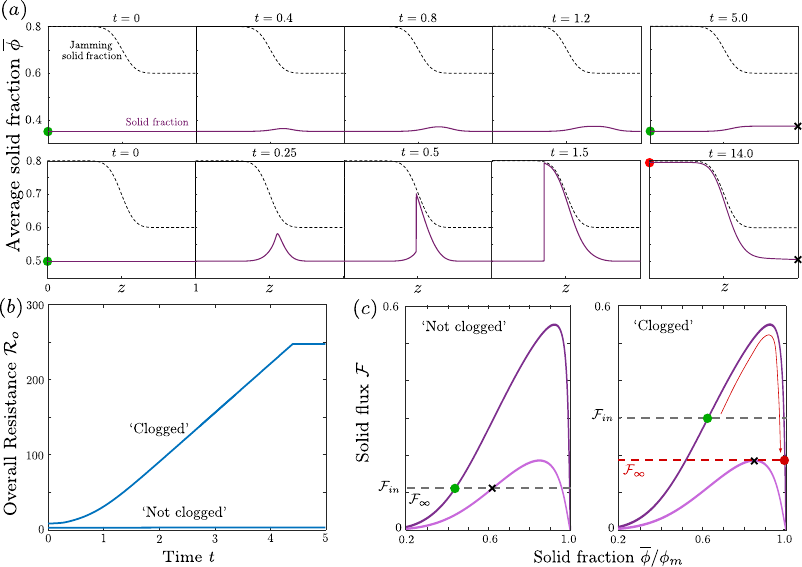}
\caption{(a) Evolving solutions for a spatially varying jamming fraction $\phi_m$ decreasing from $0.8$ to $0.6$ (dashed line) along a pipe, with $Da = 10^{-1}$, uniform radius $R=1$, and uniform initial conditions $\o{\phi}(t=0) = 0.35$ (upper row) and $\o{\phi}(t=0) = 0.5$ (lower row). (b) The associated evolution of the flow resistance in each case. (c) The solid flux in each case, comparing the flux for the highest and lowest values of $\phi_m$, with fluxes and symbols as in figure~\ref{fig:4}.}
\label{fig:5}
\end{figure}

\section{Discussion}

We have built a mechanistic continuum model that allows for spatio-temporal variations in the volume fraction of particle suspensions flowing through a pipe. A key feature of our approach is the separation of suspension flow into a ``wet solid'' phase, which tracks particles moving with the suspending fluid, and a differential Darcy phase, which tracks flow of suspending fluid past the particles; the constitutive properties of each of these phases can, in principle, be directly measured. 

In pipes with a sufficiently small constriction or low solid fraction, we find steady solutions in which a given upstream solid fraction increases as it passes through the constriction, maintaining the same solid and total flux. 
A key finding here is that significant spatio-temporal volume fraction variations can emerge spontaneously if the constriction is large enough, resulting in an abrupt transition to a high-particle-fraction, high-resistance ``clogged'' state, which develops at the entrance to the constriction and propagates back upstream. The resultant steady state in this case is controlled by the downstream constriction, rather than the upstream particle fraction, and has a particle fraction that drops, rather than increases, through the constriction (akin to observations of self-filtration; e.g. \cite{Kulkarni_2010}). 

The transition to a ``clogged'' state follows directly from the existence of a maximal solid flux at an intermediate particle fraction that depends on geometry and particle properties, and can therefore vary in space. More specifically, the emergence of ``clogging'' can occur whenever the flux in one region (e.g. a wider region of the pipe) is higher than the maximal flux in another region (e.g. a constriction; see figure~\ref{fig:3}).

The presence of this maximal solid flux emerges from our constitutive assumptions. In a suspension with fixed overall flux, without a differential Darcy flow, the flow speed of the ``wet solid'' phase would be fixed and the solid flux would linearly increase with particle fraction (the flux of interstitial fluid that moves at the same speed as the particles would reduce). However, in our model we include differential flow of fluid through the pore space of the bulk suspension, and the proportion of the overall flux taken up by each phase is determined by their relative resistances, which effectively act in parallel (figure~\ref{fig:schem}). As the solid fraction approaches its maximal jamming fraction, the resistance of the wet solid phase increases drastically, and more of the flux is pushed through the differential Darcy phase, leading to a reduction in the particle flux (figure~\ref{fig:2_II}). The emergence of the `clogged' state in a constricted channel is a consequence of the fact that the maximal solid flux decreases as the pipe radius decreases. The wet-solid resistance has a stronger dependence on radius than the resistance to Darcy flow (see \S\ref{darcy_f_sec}); Darcy flux is thus promoted for smaller radii, leading to a reduction in the attainable solid flux there.

In addition to studying constrictions, we have demonstrated that, in principle, variations in particle properties (parameterised here by variation in the jamming fraction $\phi_m$) can similarly lead to the emergence of ``clogging''. This provides a qualitative explanation for recent experimental results on blood flow from patients with sickle cell disease~\citep{Szafraniec2022,szafraniec2025suspension}, in which spatial variations in hematocrit were observed to emerge because of deoxygenation-induced stiffening of red blood cells in downstream regions of flow. Specifically, our results show that the mechanism for these results might be the same as previous results on clogging in constricted channels: the particle (or cell) flux in certain regions of the channel is higher than the maximal flux in other regions. These results will contribute to our understanding of vaso-occlusion and other pathological outcomes related to blood rheology in sickle cell disease. 

Our model is continuum and deterministic; it therefore differs in approach from stochastic and discrete theories for the effect of constrictions on particle suspensions \citep{Marin2018, Souzy2020, Zuriguel2014}, but can be interpreted as parameterising particle-scale effects mechanistically over long length and time scales. As an example, previous work on clogging has identified a probability of particles forming a temporary bridge between confining walls (``bridging''), contributing to a local build up of particles \citep{Dressaire2017, Vani2022}. To directly capture such effects, including the intermittent nature of particle bridges, our model would need to be extended to include stochasticity. However, our model can be interpreted as reflecting this process at the macroscale by relating the probability of bridging to the resistance of the suspension phase, which allows fluid to flow differentially through the Darcy phase. 

Further additional physics could be incorporated into our model for detailed quantitative comparisons with experiments on synthetic and biological suspensions. For example, allowing particle slip on the walls would allow for the prediction of an unyielded but flowing state, as has been observed experimentally \citep{Szafraniec2022,szafraniec2025suspension}. Improvements could also be made in capturing the rheology of deformable particles beyond our simple approximation of a change in the jamming fraction: particle deformability would also affect the ``wet solid'' material properties, as well as the presence of a particle-free layer at the walls. More generally, relaxing the lubrication approximation would allow the model to describe pipes with sharper along-flow variations in their properties. Our model provides a step towards a broadly applicable continuum framework that links particle motion and macroscopic material properties during self-filtration and clogging of suspensions.

\vspace{.5cm}

We would like to acknowledge numerous helpful discussions with Valeria Ciccone, Anne Juel, Hannah Szfraniec, David Wood, Parisa Bazazi, John Higgins, Timm Kr\"uger and Howard Stone.  Declaration of interests. The authors report no conflict of interest. P.P. is supported by a UKRI Future Leaders Fellowship [MR/V022385/1].

%\appendix
\begin{appendices}

\section{Cross-sectional averaging}\label{Appendix1}

\subsection{Radial integration}

\begin{figure}
\centering
\includegraphics[width=0.5\linewidth]{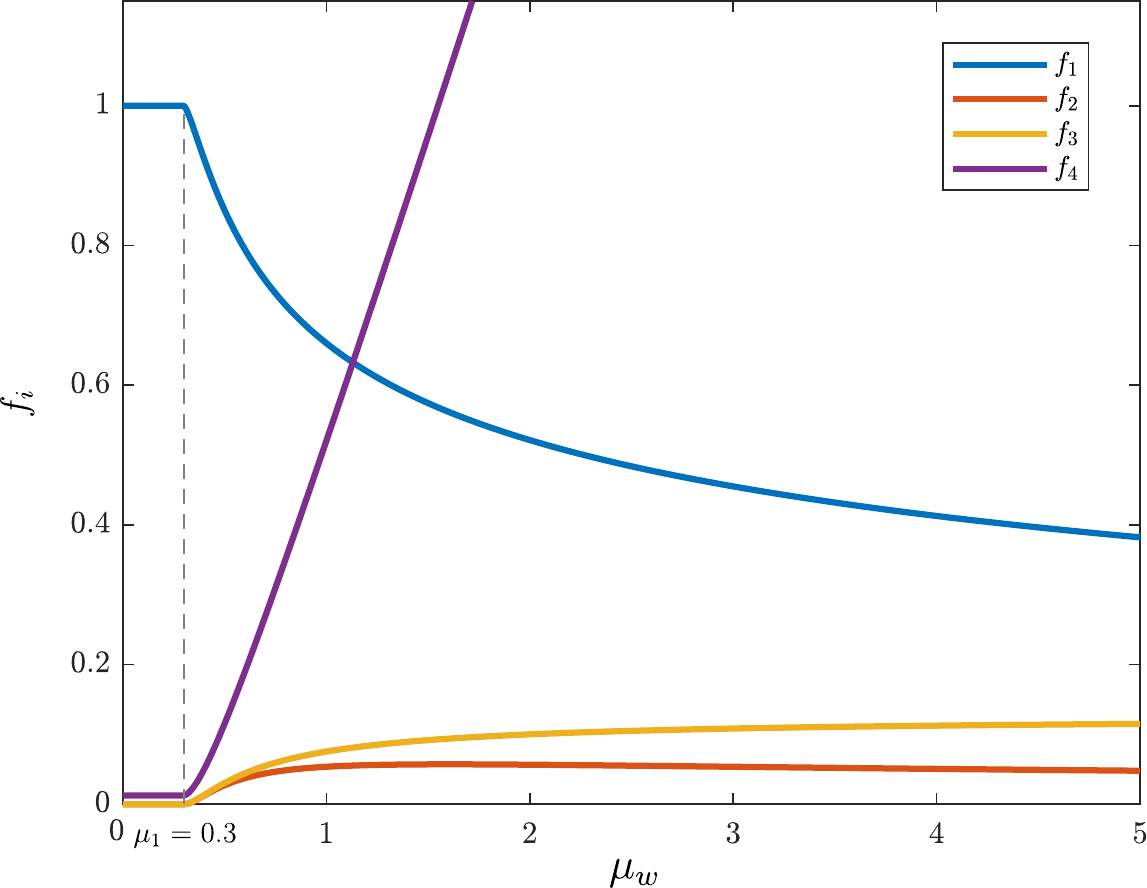}
 \caption{The average functions $f_i$, defined in (\ref{phi_o})--(\ref{us_uf_avg}), which depend on the grouping $\mu_w= GR/2p_s$, and represent the scaled average solid fraction, solid flux, solid velocity and Darcy velocity, respectively. For $\mu_w < \mu_1$, the entire pipe is clogged with $\phi = \phi_m$ everywhere and $u_s = 0$ Here, we have taken $Da = 10^{-1}$ and $\phi_m = 0.8$.}
\label{fig:functions}
\end{figure}

To solve (\ref{av2}) we need to compute various radial averages, which follow from suitable integration of the expressions for $\phi$ (\ref{phiprofile}) and $u_s$ (\ref{velprof}). We can isolate the dependence on the grouping $\mu_w = GR/2p_s$ in these expression to write the relevant averages in terms of four dimensionless functions,
\begin{align}
    \frac{\o{\phi}}{\phi_m} &= f_1(\mu_w), & \f{\o{ u_s\phi}}{G\phi_m R^2} &= f_2(\mu_w),\label{phi_o}\\
    \frac{\o{u_s}}{G R^2} &= f_3(\mu_w), & 
    \frac{\o{u_D}}{Da \, G} &=f_4(\mu_w, \phi_m),\label{us_uf_avg}
\end{align}
where the functions $f_i$ can be computed analytically by integrating the expressions in (\ref{phiprofile}), (\ref{velprof}) and the final equation in (\ref{lub1}). The expressions are rather analytically involved, and are listed below for reference. Note that $f_4$ is really just the radially averaged permeability function, coming from Darcy's law (\ref{lub1}), and thus depends in general on both $\mu_w$ and $\phi_m$. 

Figure \ref{fig:functions} shows how these functions vary with $\mu_w$. For $\mu_w<\mu_1$, the entire pipe is clogged: $\phi = \phi_m$ everywhere and $u_s = 0$; hence $f_1 =1$ and $f_2 = f_3 = 0$ over that range. As $\mu_w$ is increased beyond this critical value, the material starts to flow, so $f_1$ decreases and the other functions increase. 

In terms of these functions, the governing equations (\ref{av2}) become
\begin{equation}
    \pder{\phi_m f_1}{t} + \f{1}{R^2}\pder{}{z}(R^4\phi_m G f_2) = 0,
    \qquad
     \f{Q}{\pi R^2} = R^2Gf_3 + Da \, G f_4.
     \label{ap_av2}
\end{equation}
The latter expression can be inverted to give an explicit expression for the pressure gradient $G$, 
\begin{gather}
    G(\mu_w, R, \phi_m) = \f{Q}{\pi R^4(f_3 + Da f_4/R^2)},\label{ap_G}
\end{gather}
and so (\ref{ap_av2}) becomes 
\begin{equation}
    \pder{}{t}(\phi_m f_1) + \f{1}{R^2}\pder{ \mathcal{F}}{z} = 0, 
    \qquad
    \mathcal{F}(\mu_w,\phi_m, R) \equiv \f{\phi_m f_2 Q}{\pi \l(f_3 + Da f_4/R^2\r)},
    \label{flux_AP}
\end{equation}
where $\mathcal{F}  = R^2\o{u_s \phi}$ is the solid flux. 

The total pressure drop along the channel can be calculated from 
\begin{gather}
  \Delta p = \int_{0}^{1} G  \, dz =
 Q/\pi\int_{0}^{1} \l(R^4(f_3 + Da f_4/R^2)\r)^{-1}  dz, %\label{Q_DelP}
\end{gather}
which also determines the overall resistance %(\ref{resis}) 
$\mathcal{R}_o = \Delta p/Q = 1/\pi\int_0^1 \l(R^2(f_3 + Da f_4/R^2)\r)^{-1}  dz $ and the individual resistances of each phase,
\be
\mathcal{R}_s = \frac{\Delta p}{\pi R^2 \o{u_s}} = \l(\frac{\Delta p}{Q} \r) \frac{f_3 + Da f_4/R^2}{f_3},
\qquad
\mathcal{R}_D = \frac{\Delta p}{\pi R^2 \o{u_D}} = \l(\frac{\Delta p}{Q}\r) \frac{f_3 + Da f_4/R^2}{Da f_4/R^2}.
\ee

Note that the radius $R$ enters the expression for the solid flux $\mathcal{F}$ in (\ref{flux_AP}) only in the term multiplying the Darcy number. This feature illustrates how the reduction in the flux for smaller pipe radii, which controls the emergent ``clogging'' behaviour that we observe here, is physically a result of the fact that the Darcy flow is relatively larger when the radius is reduced, as discussed in \S~\ref{darcy_f_sec}. Relatively speaking, more of the total flux is taken up by differential seepage flow in a thinner pipe.

\subsection{Integral functions}
Explicit forms of the integral functions defined above are listed below for reference:
\begin{gather}
\begin{aligned}
f_1 &=2(\mu_w(-3 + 2(\mu_w - \mu_1)^{1/2}) + 3\mu_1 + 2(\mu_w - \mu_1)^{1/2}(3 + 2\mu_1)\\
&- 6(1 + \mu_1)log(1 + (\mu_w - \mu_1)^{1/2})/(3\mu_w^2) + (\mu_1/\mu_w)^2\\\\
 f_2 &= (2\mu_1 - \mu_w)/(2\mu_w^2) - (\mu_w - \mu_1)^{1/2}/(7\mu_w) - (\mu_1^2 - 1)/(2\mu_w^3) - \mu_1/(2\mu_w^2) + 1/(6\mu_w) \\
 &+ (\mu_1 + 1)/(4\mu_w^2) - \mu_1^3/(6\mu_w^4) - ((\mu_w - \mu_1)^{1/2}(- 64\mu_1^3 + 140\mu_1^2\mu_w - 119\mu_1^2\\
 &- 70\mu_1\mu_w^2 + 210\mu_1\mu_w + 70\mu_1 - 105\mu_w^2 +105))/(105\mu_w^4) + ((8\mu_1 - 7)(\mu_w - \mu_1)^{1/2})/(35\mu_w^2)\\
 &- (\mu_1(- \mu_1^2 + 2\mu_1\mu_w - \mu_w^2 + 1))/(2\mu_w^4)+ (\mu_1^2(\mu_1 - 1))/(4\mu_w^4) \\
 &+ ((\mu_w - \mu_1)^{1/2}(32\mu_1^2 - 70\mu_1\mu_w + 7\mu_1 + 35\mu_w^2 - 35))/(105\mu_w^3) + (\mu_1^2(\mu_1 - \mu_w)^2)/(4\mu_w^4)\\
 &+ (log((\mu_w - \mu_1 + 2(\mu_w - \mu_1)^{1/2} + 1))(\mu_1 + 1)(- \mu_1^2 + 2\mu_1\mu_w - \mu_w^2 + 1))/(2\mu_w^4) \\\\
f_3 &= (\l(\mu_1/\mu_w\r)^2(\mu_1/\mu_w - 1/2))/2 - \mu_1/6\mu_w  + (\l(\mu_1/\mu_w\r)^2(\mu_1/\mu_w - 1)^2)/4 - (5\l(\mu_1/\mu_w\r)^4)/24\\
&+ 1/8\\\\
f_4 &=-(16\mu_1^2(\mu_w - \mu_1)^{1/2} - 24\mu_w^2(\mu_w - \mu_1)^{1/2} + 60\phi_m^3(\mu_w - \mu_1)^{1/2} + 15\mu_1\mu_w^2 + 30\mu_1\phi_m^3\\
&+ 45\mu_w^2\phi_m - 30\mu_w\phi_m^3 - 5\mu_1^3 - 15\mu_w^2 - 10\mu_w^3 - 30\phi_m^3log((\mu_w - \mu_1 + 2(\mu_w - \mu_1)^{1/2} + 1))\\
&+ 15\mu_1^2\phi_m^3 - 45\mu_w^2\phi_m^2 + 8\mu_1\mu_w(\mu_w - \mu_1)^{1/2}- 30\mu_1\phi_m^3log((\mu_w - \mu_1 + 2(\mu_w - \mu_1)^{1/2} + 1))\\
&- 24\mu_1^2\phi_m(\mu_w - \mu_1)^{1/2}+ 40\mu_1\phi_m^3(\mu_w - \mu_1)^{1/2} + 36\mu_w^2\phi_m(\mu_w - \mu_1)^{1/2} + 20\mu_w\phi_m^3(\mu_w - \mu_1)^{1/2}\\
&- 12\mu_1 \mu_w\phi_m(\mu_w - \mu_1)^{1/2})/(15\mu_w^2\phi_m^2)\\
 \end{aligned}
\end{gather}

\section{Time-dependent numerical method}\label{AppendixNum}

 Equation \eqref{av2} is a quasi-linear hyperbolic partial differential equation which can give rise to shocks for certain initial conditions. To handle these, we use the finite-volume package Clawpack \citep{clawpack}. Clawpack uses a high-resolution Godunov-type method, which employs Riemann solvers at cell interfaces to capture the propagation of waves accurately. We use second-order Godunov method with a van Leer flux limiter to solve the Riemann problems for the whole domain.

For the boundary conditions, we use zero-order extrapolation by implementing ghost cells that extend the cell values from the upstream and downstream boundaries; these are updated at every time step. This ensures that no spurious waves are generated at the boundaries that might alter the solution. The system of equations that we input in Clawpack are in conservative form to ensure accuracy. Using definitions of integral functions from \eqref{us_uf_avg}, we define a new variable $q$ for computational purposes,
\begin{align}
    q &= R^2\phi_m f_1(\mu_w).
\end{align}
We encode a spatially varying radius $R(z$) or maximum packing fraction $\phi_m(z)$ through two extra equations in combination with the explicit form of the conservation equation \eqref{av2},
\begin{equation}
        \pder{q}{t} + \pder{}{z}\l(\frac{\phi_m f_2}{f_3 + Da f_4/R^2  }\r)=0,
        \qquad
        \pder{R}{t} = 0,
        \qquad
        \pder{\phi_m}{t} = 0.
        \label{q_eq}
\end{equation}
The specific form of $R(z)$ or $\phi_m(z)$ is then set as an `initial' condition for that variable. The form for radial constrictions was given in (\ref{radial_z}), and for simulations with varying jamming fraction we set  $   \phi_m(z,t=0) =  a - (a-b)\left\{1 + erf[(z-0.5)10]\right\}/2$, where $a$ and $b$ represent the upstream and downstream values. 

\begin{figure}
\centering
\includegraphics[width = \linewidth]{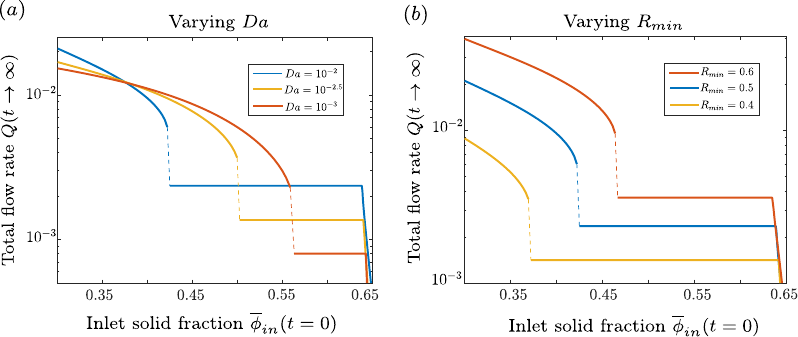}
\caption{
Steady results for fixed pressure drop, showing how the final steady-state total flux $Q$ decreases with the initial inlet solid fraction, for $(a)$ varying $Da$ with $R_{min}= 0.5$ and $(b)$ varying constriction ratio $R_{min}$ with $Da = 10^{-2}$. Here $\phi_m = 0.65$. 
}
\label{fig:Q_new}
\end{figure}

\section{Fixed pressure drop}
\label{ap_press}

The time-dependent numerical method would become significantly more numerically expensive if the pressure drop along the pipe, as opposed to the total flux, was fixed (this would require iteration of an integral constraint on the flux at each time step). However, the steady solutions can be straightforwardly inverted to give the case with a fixed pressure drop: results follow directly from the overall resistance in figure~\ref{fig:Resist_new}(b). The resistance provides a measure of the pressure drop required to sustain a given total flux, and so its inverse provides the flux required to sustain a given pressure drop in steady state. This quantity is plotted in figure~\ref{fig:Q_new}, for various different parameters, which shows how the total flux drops discontinuously upon entering the ``clogged'' regime. 

\end{appendices}

\bibliographystyle{jfm}
\bibliography{jfm}

\begin{thebibliography}{34}
\expandafter\ifx\csname natexlab\endcsname\relax\def\natexlab#1{#1}\fi
\def\au#1{#1} \def\ed#1{#1} \def\yr#1{#1}\def\at#1{#1}\def\jt#1{\textit{#1}}
  \def\bt#1{#1}\def\bvol#1{\textbf{#1}} \def\vol#1{#1} \def\pg#1{#1}
  \def\publ#1{#1}\def\arxiv#1{#1}\def\org#1{#1}\def\st#1{\textit{#1}}

\bibitem[Ahnert {\em et~al.\/}(2019)Ahnert, Münch \& Wagner]{Ahnert2019}
{\sc \au{Ahnert, T.}, \au{Münch, A.} \& \au{Wagner, B.}} \yr{2019}  \at{Models
  for the two-phase flow of concentrated suspensions}.  \jt{Eur. J. Appl.
  Math.}  \bvol{30},  \pg{557--584}.

\bibitem[Barker {\em et~al.\/}(2015)Barker, Schaeffer, Bohorquez \&
  Gray]{Barker_2015}
{\sc \au{Barker, T.}, \au{Schaeffer, D.~G.}, \au{Bohorquez, P.} \& \au{Gray, J.
  M. N.~T.}} \yr{2015}  \at{Well-posed and ill-posed behaviour of the
  ${\it\mu}(i)$-rheology for granular flow}.  \jt{J. Fluid Mech.}  \bvol{779},
  \pg{794–818}.

\bibitem[Barnes(1995)]{barnes1995review}
{\sc \au{Barnes, H~A}} \yr{1995}  \at{A review of the slip (wall depletion) of
  polymer solutions, emulsions and particle suspensions in viscometers: its
  cause, character, and cure}.  \jt{J. Non-Newton. Fluid Mech.}  \bvol{56}~(3),
   \pg{221--251}.

\bibitem[Baumgarten \& Kamrin(2019)]{Baumgarten2019}
{\sc \au{Baumgarten, A.~S.} \& \au{Kamrin, K.}} \yr{2019}  \at{A general
  fluid-sediment mixture model and constitutive theory validated in many flow
  regimes}.  \jt{J. Fluid Mech.}  \bvol{861},  \pg{721--764}.

\bibitem[Boyer {\em et~al.\/}(2011)Boyer, Guazzelli \& Pouliquen]{Boyer2011}
{\sc \au{Boyer, F.}, \au{Guazzelli, É.} \& \au{Pouliquen, O.}} \yr{2011}
  \at{Unifying suspension and granular rheology}.  \jt{Phys. Rev. Lett.}
  \bvol{107}.

\bibitem[Bächer {\em et~al.\/}(2017)Bächer, Schrack \& Gekle]{Bacher2017}
{\sc \au{Bächer, C.}, \au{Schrack, L.} \& \au{Gekle, S.}} \yr{2017}
  \at{Clustering of microscopic particles in constricted blood flow}.
  \jt{Phys. Rev. Fluids}  \bvol{2}.

\bibitem[{Clawpack Development Team}(2024)]{clawpack}
{\sc \au{{Clawpack Development Team}}} \yr{2024} Clawpack software. Version
  5.10.0.

\bibitem[Cohen \& Mahadevan(2013)]{Cohen2013}
{\sc \au{Cohen, S. I.~A.} \& \au{Mahadevan, L.}} \yr{2013}  \at{Hydrodynamics
  of hemostasis in sickle-cell disease}.  \jt{Phys. Rev. Lett.}  \bvol{110}.

\bibitem[Dagois-Bohy {\em et~al.\/}(2015)Dagois-Bohy, Hormozi, Guazzelli \&
  Pouliquen]{Dagois-Bohy_2015}
{\sc \au{Dagois-Bohy, S.}, \au{Hormozi, S.}, \au{Guazzelli, É.} \&
  \au{Pouliquen, O.}} \yr{2015}  \at{Rheology of dense suspensions of
  non-colloidal spheres in yield-stress fluids}.  \jt{J. Fluid Mech.}
  \bvol{776},  \pg{R2}.

\bibitem[Dbouk {\em et~al.\/}(2013)Dbouk, Lobry \& Lemaire]{Dbouk_2013}
{\sc \au{Dbouk, T.}, \au{Lobry, L.} \& \au{Lemaire, E.}} \yr{2013}  \at{Normal
  stresses in concentrated non-brownian suspensions}.  \jt{J. Fluid Mech.}
  \bvol{715},  \pg{239–272}.

\bibitem[Dincau {\em et~al.\/}(2023)Dincau, Dressaire \& Sauret]{Dincau2023}
{\sc \au{Dincau, B.}, \au{Dressaire, E.} \& \au{Sauret, A.}} \yr{2023}
  \at{Clogging: The self-sabotage of suspensions}.  \jt{Phys. Today}
  \bvol{76},  \pg{24--30}.

\bibitem[Dressaire \& Sauret(2017)]{Dressaire2017}
{\sc \au{Dressaire, E.} \& \au{Sauret, A.}} \yr{2017}  \at{Clogging of
  microfluidic systems}.  \jt{Soft Matter}  \bvol{13},  \pg{37--48}.

\bibitem[Farutin {\em et~al.\/}(2018)Farutin, Shen, Prado, Audemar,
  Ez-Zahraouy, Benyoussef, Polack, Harting, Vlahovska, Podgorski, Coupier \&
  Misbah]{Farutin2018}
{\sc \au{Farutin, A.}, \au{Shen, Z.}, \au{Prado, G.}, \au{Audemar, V.},
  \au{Ez-Zahraouy, H.}, \au{Benyoussef, A.}, \au{Polack, B.}, \au{Harting, J.},
  \au{Vlahovska, P.~M.}, \au{Podgorski, T.}, \au{Coupier, G.} \& \au{Misbah,
  C.}} \yr{2018}  \at{Optimal cell transport in straight channels and
  networks}.  \jt{Phys. Rev. Fluids}  \bvol{3}.

\bibitem[Gallier {\em et~al.\/}(2014)Gallier, Lemaire, Peters \&
  Lobry]{Gallier_2014}
{\sc \au{Gallier, S.}, \au{Lemaire, E.}, \au{Peters, F.} \& \au{Lobry, L.}}
  \yr{2014}  \at{Rheology of sheared suspensions of rough frictional
  particles}.  \jt{J. Fluid Mech.}  \bvol{757},  \pg{514–549}.

\bibitem[Guazzelli \& Pouliquen(2018)]{Guazzelli2018}
{\sc \au{Guazzelli, E.} \& \au{Pouliquen, O.}} \yr{2018}  \at{Rheology of dense
  granular suspensions}.  \jt{J. Fluid Mech.}  \bvol{852}.

\bibitem[Haw(2004)]{Haw_2004}
{\sc \au{Haw, M.~D.}} \yr{2004}  \at{Jamming, two-fluid behavior, and
  ``self-filtration'' in concentrated particulate suspensions}.  \jt{Phys. Rev.
  Lett.}  \bvol{92},  \pg{185506}.

\bibitem[Koh {\em et~al.\/}(1994)Koh, Hookham \& Leal]{Koh1994}
{\sc \au{Koh, C.~J.}, \au{Hookham, P.} \& \au{Leal, L.~G.}} \yr{1994}  \at{An
  experimental investigation of concentrated suspension flows in a rectangular
  channel}.  \jt{J. Fluid Mech.}  \bvol{266},  \pg{1--32}.

\bibitem[Kulkarni {\em et~al.\/}(2010)Kulkarni, Metzger \&
  Morris]{Kulkarni_2010}
{\sc \au{Kulkarni, S.~D.}, \au{Metzger, B.} \& \au{Morris, J.~F.}} \yr{2010}
  \at{Particle-pressure-induced self-filtration in concentrated suspensions}.
  \jt{Phys. Rev. E}  \bvol{82},  \pg{010402}.

\bibitem[Lyon \& Leal(1998)]{Lyon1998}
{\sc \au{Lyon, M.~K.} \& \au{Leal, L.~G.}} \yr{1998}  \at{An experimental study
  of the motion of concentrated suspensions in two-dimensional channel flow.
  part 1. monodisperse systems}.  \jt{J. Fluid Mech.}  \bvol{363},
  \pg{25--56}.

\bibitem[Marin {\em et~al.\/}(2018)Marin, Lhuissier, Rossi \&
  Kähler]{Marin2018}
{\sc \au{Marin, A.}, \au{Lhuissier, H.}, \au{Rossi, M.} \& \au{Kähler, C.J.}}
  \yr{2018}  \at{Clogging in constricted suspension flows}.  \jt{Phys. Rev. E}
  \bvol{97}.

\bibitem[Marin \& Souzy(2024)]{Marin2024}
{\sc \au{Marin, A.} \& \au{Souzy, M.}} \yr{2024}  \at{Clogging of noncohesive
  suspension flows}.  \jt{Annu. Rev. Fluid Mech.}  \bvol{57},  \pg{89--116}.

\bibitem[Mondal {\em et~al.\/}(2016)Mondal, Wu \& Sharma]{Mondal2016}
{\sc \au{Mondal, S.}, \au{Wu, C.~H.} \& \au{Sharma, M.~M.}} \yr{2016}
  \at{Coupled {CFD-DEM} simulation of hydrodynamic bridging at constrictions}.
  \jt{Int. J. Multiph. Flow}  \bvol{84},  \pg{245--263}.

\bibitem[Parry \& Millet(2010)]{Parry2010}
{\sc \au{Parry, A.~J.} \& \au{Millet, O.}} \yr{2010}  \at{Modeling blockage of
  particles in conduit constrictions: Dense granular-suspension flow}.  \jt{J.
  Fluids Eng.}  \bvol{132},  \pg{0113021--01130210}.

\bibitem[Phillips {\em et~al.\/}(1992)Phillips, Armstrong, Brown, Graham \&
  Abbott]{Phillips1992}
{\sc \au{Phillips, R.~J.}, \au{Armstrong, R.~C.}, \au{Brown, R.~A.},
  \au{Graham, A.~L.} \& \au{Abbott, J.~R.}} \yr{1992}  \at{A constitutive
  equation for concentrated suspensions that accounts for shear-induced
  particle migration}.  \jt{Phys. Fluids A}  \bvol{4},  \pg{30--40}.

\bibitem[Souzy {\em et~al.\/}(2020)Souzy, Zuriguel \& Marin]{Souzy2020}
{\sc \au{Souzy, M.}, \au{Zuriguel, I.} \& \au{Marin, A.}} \yr{2020}
  \at{Transition from clogging to continuous flow in constricted particle
  suspensions}.  \jt{Phys. Rev. E}  \bvol{101}.

\bibitem[Stathoulopoulos {\em et~al.\/}(2025)Stathoulopoulos, König,
  Ramachandran \& Balabani]{Stavroula_2025}
{\sc \au{Stathoulopoulos, A.}, \au{König, C.~S.}, \au{Ramachandran, S.} \&
  \au{Balabani, S.}} \yr{2025}  \at{Statin-treated {RBC} dynamics in a
  microfluidic porous-like network}.  \jt{Microvasc. Res.}  \bvol{158},
  \pg{104765}.

\bibitem[Stickel \& Powell(2005)]{Stickel2005}
{\sc \au{Stickel, Jonathan~J.} \& \au{Powell, Robert~L.}} \yr{2005}  \at{Fluid
  mechanics and rheology of dense suspensions}.  \jt{Annu. Rev. Fluid Mech.}
  \bvol{37},  \pg{129--149}.

\bibitem[Szafraniec {\em et~al.\/}(2025)Szafraniec, Bull, Higgins, Stone,
  Kr{\"u}ger, Pearce \& Wood]{szafraniec2025suspension}
{\sc \au{Szafraniec, H.~M.}, \au{Bull, F.}, \au{Higgins, J.~M.}, \au{Stone,
  H.~A.}, \au{Kr{\"u}ger, T.}, \au{Pearce, P.} \& \au{Wood, D.~K.}} \yr{2025}
  \at{Suspension physics govern the multiscale dynamics of blood flow in sickle
  cell disease}.  \jt{bioRxiv}  \pg{p. 2025.03.13.642599}.

\bibitem[Szafraniec {\em et~al.\/}(2022)Szafraniec, Valdez, Iffrig, Lam,
  Higgins, Pearce \& Wood]{Szafraniec2022}
{\sc \au{Szafraniec, H.~M.}, \au{Valdez, J.~M.}, \au{Iffrig, E.}, \au{Lam,
  W.~A.}, \au{Higgins, J.~M.}, \au{Pearce, P.} \& \au{Wood, D.~K.}} \yr{2022}
  \at{Feature tracking microfluidic analysis reveals differential roles of
  viscosity and friction in sickle cell blood}.  \jt{Lab Chip}  \bvol{22},
  \pg{1565--1575}.

\bibitem[Tapia {\em et~al.\/}(2024)Tapia, Hong, Aussillous \&
  Guazzelli]{tapia2024rheology}
{\sc \au{Tapia, F.}, \au{Hong, C.}, \au{Aussillous, P.} \& \au{Guazzelli,
  {\'E}.}} \yr{2024}  \at{Rheology of suspensions of non-brownian soft spheres
  across the jamming and viscous-to-inertial transitions}.  \jt{Phys. Rev.
  Lett.}  \bvol{133}~(8),  \pg{088201}.

\bibitem[Vani {\em et~al.\/}(2022)Vani, Escudier \& Sauret]{Vani2022}
{\sc \au{Vani, N.}, \au{Escudier, S.} \& \au{Sauret, A.}} \yr{2022}
  \at{Influence of the solid fraction on the clogging by bridging of
  suspensions in constricted channels}.  \jt{Soft Matter}  \bvol{18}~(36),
  \pg{6987–6997}.

\bibitem[Wood {\em et~al.\/}(2012)Wood, Soriano, Mahadevan, Higgins \&
  Bhatia]{Dave2012}
{\sc \au{Wood, D.~K.}, \au{Soriano, A.}, \au{Mahadevan, L.}, \au{Higgins,
  J.~M.} \& \au{Bhatia, S.~N.}} \yr{2012}  \at{A biophysical indicator of
  vaso-occlusive risk in sickle cell disease}.  \jt{Sci. Transl. Med.}
  \bvol{4}~(123),  \pg{123ra26--123ra26}.

\bibitem[Zarraga {\em et~al.\/}(2000)Zarraga, Hill \& Leighton]{Zarraga2000}
{\sc \au{Zarraga, I.~E.}, \au{Hill, D.~A.} \& \au{Leighton, D.~T.}} \yr{2000}
  \at{The characterization of the total stress of concentrated suspensions of
  noncolloidal spheres in newtonian fluids}.  \jt{J. Reol.}  \bvol{44},
  \pg{185--220}.

\bibitem[Zuriguel {\em et~al.\/}(2014)Zuriguel, Parisi, Hidalgo, Lozano, Janda,
  Gago, Peralta, Ferrer, Pugnaloni, Clément, Maza, Pagonabarraga \&
  Garcimartín]{Zuriguel2014}
{\sc \au{Zuriguel, I.}, \au{Parisi, D.~R.}, \au{Hidalgo, R.~C.}, \au{Lozano,
  C.}, \au{Janda, A.}, \au{Gago, P.~A.}, \au{Peralta, J.~P.}, \au{Ferrer,
  L.~M.}, \au{Pugnaloni, L.~A.}, \au{Clément, E.}, \au{Maza, D.},
  \au{Pagonabarraga, I.} \& \au{Garcimartín, A.}} \yr{2014}  \at{Clogging
  transition of many-particle systems flowing through bottlenecks}.  \jt{Sci.
  Rep.}  \bvol{4}.

\end{thebibliography}

\end{document}